\documentclass[aps,prl,reprint,groupedaddress,floatfix,nobalancelastpage]{revtex4-2}

\usepackage{amsmath,amssymb,amsthm,dcolumn}
\usepackage{graphicx}
\usepackage{xcolor}
\usepackage{hyperref}
\hypersetup{colorlinks,linkcolor={blue!90!black},citecolor={blue!90!black},urlcolor={blue!90!black}}
\usepackage{stmaryrd}
\usepackage{enumitem}
\usepackage{accents}
\usepackage{twemojis}

\newcommand{\figref}[2]{[Fig.~\hyperref[#1]{\ref*{#1}(#2)}]}
\newcommand{\figrefi}[2]{[Fig.~\hyperref[#1]{\ref*{#1}(#2)}, inset]}
\newcommand{\textfigref}[2]{Fig.~\hyperref[#1]{\ref*{#1}(#2)}}
\newcommand{\textfigureref}[2]{Figure~\hyperref[#1]{\ref*{#1}(#2)}}
\newcommand{\wholefigref}[1]{(Fig.~\ref{#1})}

\newcommand{\textwholefigref}[1]{Fig.~\ref{#1}}

\newcommand{\figrefp}[2]{\hyperref[#1]{\ref*{#1}(#2)}}

\DeclareMathAlphabet{\mathcal}{OMS}{cmsy}{m}{n}
\DeclareMathAlphabet{\mathbfsf}{\encodingdefault}{\sfdefault}{b}{n}

\definecolor{linkcolor}{HTML}{223096}
\hypersetup{colorlinks,allcolors=linkcolor}

\bibpunct{\textcolor{linkcolor}{[}}{\textcolor{linkcolor}{]}}{\textcolor{linkcolor}{,}}{n}{}{;}
\renewcommand{\eqref}[1]{\hyperref[#1]{(\ref*{#1})}}

\renewcommand{\ge}{\geqslant}
\newcommand{\Pii}{\mathit{\Pi}}

\begin{document}
\title{Mechanics of poking a cyst}
\author{Shiheng Zhao}
\affiliation{Max Planck Institute for the Physics of Complex Systems, N\"othnitzer Straße 38, 01187 Dresden, Germany}
\affiliation{Center for Systems Biology Dresden, Pfotenhauerstraße 108, 01307 Dresden, Germany}
\affiliation{\mbox{Max Planck Institute of Molecular Cell Biology and Genetics, Pfotenhauerstraße 108, 01307 Dresden, Germany}}
\author{Pierre A. Haas}
\email{haas@pks.mpg.de}
\affiliation{Max Planck Institute for the Physics of Complex Systems, N\"othnitzer Straße 38, 01187 Dresden, Germany}
\affiliation{Center for Systems Biology Dresden, Pfotenhauerstraße 108, 01307 Dresden, Germany}
\affiliation{\mbox{Max Planck Institute of Molecular Cell Biology and Genetics, Pfotenhauerstraße 108, 01307 Dresden, Germany}}

\date{\today}

\begin{abstract}
Indentation tests are classical tools to determine material properties. For biological samples such as cysts of cells, however, the observed force-displacement relation cannot be expected to follow predictions for simple materials. Here, by solving the Pogorelov problem of a point force indenting an elastic shell for a purely nonlinear material, we discover that complex material behaviour can even give rise to new scaling exponents in this force-displacement relation. In finite-element simulations, we show that these exponents are surprisingly robust, persisting even for thick shells indented with a sphere. By scaling arguments, we generalise our results to pressurised and pre-stressed shells, uncovering additional new scaling exponents. We find these predicted scaling exponents in the force-displacement relation observed in cyst indentation experiments. Our results thus form the basis for inferring the mechanisms that set the mechanical properties of these biological materials.
\end{abstract}

\maketitle
\renewcommand{\floatpagefraction}{.999}

Indentation tests are perhaps the most direct way of probing material properties across scales and applications~\cite{bishop45,Arunkumar2018}: Just as one is wont to poke the fruity wares peddled in supermarkets to evaluate the immediacy of their comestibility~\cite{box20}, indentation of biological samples~\cite{argatov2018indentation,jin2022} reveals mechanical properties that are intrinsically linked to their biological function~\cite{fletcher2010,Joshua2011PRL,Andreas2020,pereztirado24}. In all cases, the relation between the indentation force $F$ and the displacement $e$ of the indenter must be fitted to a mechanical model of the system to infer material properties. The correct mechanics remain elusive, however, for many biological systems.

\begin{figure}[b]
    \centering
    \includegraphics[width=0.8\linewidth]{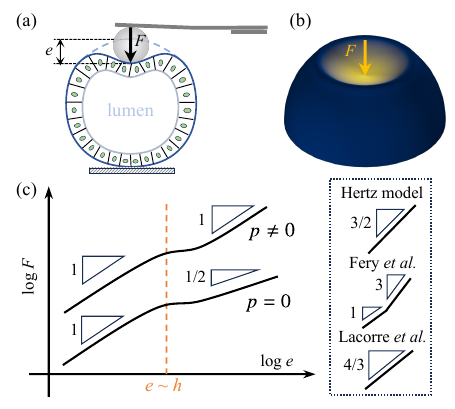}
    \caption{Mechanics of poking a cyst. (a)~A cyst consists of a spherical monolayer of cells surrounding a fluid-filled lumen. Indentation of the cyst by a microbead attached to an AFM cantilever~\cite{Shen2017} yields the relation between the poking force $F$ and the indentation $e$ of the cyst. (b)~The Pogorelov problem~\cite{pogorelov}: A thin hemispherical shell of radius $R$ and thickness $h\ll R$ is indented by a point force $F$. (c)~Scaling exponents in the force displacement relation for the unpressurised Pogorelov problem ($p=0$)~\cite{pogorelov,landaulifshitz,audoly} and the pressurised Pogorelov problem (${p\neq 0}$)~\cite{Vella2012,Vella2012PRL}. Insets: alternative scaling exponents predicted by the Hertzian contact model~\cite{Hertz,landaulifshitz,audoly} and the models of Fery \emph{et al.}~\cite{Fery2004} and Lacorre~\emph{et al.}~\cite{Lacorre2024}.}
    \label{fig1}
\end{figure}

Cysts, spherical monolayers of polarised cells surrounding a fluid-filled lumen~\cite{bovyn24}, are paradigmatic examplars of this problem: Indentation of cysts~\cite{Shen2017} by a microbead attached to an AFM cantilever~\figref{fig1}{a} quantifies their force-displacement relation, but the mechanical laws that govern it are unclear. Since cell cortices respond elastically to external forces~\cite{salbreux12,bovyn24} on the short timescales of indentation experiments, the classical Pogorelov problem~\cite{pogorelov} of a point force on a thin elastic shell~\figref{fig1}{b} provides a possible mechanical model. Pogorelov's calculation~\cite{pogorelov,landaulifshitz,audoly,gomez16} for an unpressurised shell predicts $F\sim e$ for $e\ll h$, where $h$ is the shell thickness, and $F\sim e^{1/2}$ for $e\gg h$~\figref{fig1}{c}. If instead the shell is pressurised~\cite{Vella2012,Vella2012PRL}, $F\sim e$ also holds for $e\gg h$. Meanwhile, the Hertzian contact problem of two elastic spheres~\cite{Hertz,landaulifshitz,audoly} provides an alternative mechanical model for the cyst indentation experiments, but predicts a different scaling exponent $F\sim e^{3/2}$~\figrefi{fig1}{c}. Another model imposes lumen incompressibility~\cite{Fery2004} to find the scaling $F\sim e^3$, while very recent work~\cite{Lacorre2024} predicts yet another scaling, $F\sim e^{4/3}$.   

These different models do not thus even agree on the scaling exponents (let alone on the corresponding prefactors that include the mechanical parameters to be fitted for). Additionally, the effect of pressure~\figref{fig1}{c} stresses the possible importance of different competing effects. For example, the models above do not include the pre-stress resulting from active cell contractility or cell-cell adhesion~\cite{lecuit07,lecuit11,salbreux12} that has been added in a recent membrane model~\cite{Andreas2020}. Moreover, biological materials are very soft: The elastic modulus of biological tissues including brain matter~\cite{Destrade2015,budday17} is much lower than that of inert soft materials such as rubber or hydrogels~\cite{miserez15,sheiko19,lou23}. Material nonlinearities are therefore expect to matter for the large deformations in cyst indentation experiments, but are completely absent from these models.

Here, we therefore start by solving the nonlinear Pogorelov problem: A point force indents a purely nonlinearly elastic shell. Our asymptotic calculations for thin shells reveal new scaling exponents. In finite-element simulations, we verify these results and show that, strikingly, they also apply in the experimentally relevant setting of a thick shell indented with a sphere. We then develop scaling arguments that extend these results to pressurised and pre-stressed shells and predict additional new scaling exponents. We thus classify all possible diagrams of scaling exponents. We conclude by reanalysing experimental data on cyst indentation experiments, in which we discover signatures of these scaling exponents.

Our starting point is a general incompressible polynomial hyperelastic energy density~\cite{rivlin1951,ogden,dervaux09},
\begin{align}
W=\sum_{m=0}^\infty{\sum_{n=0}^\infty{C_{mn}}(\mathcal{I}_1-3)^m(\mathcal{I}_2-3)^n},\label{eq:W}
\end{align}
expressed in terms of the principal invariants $\mathcal{I}_1,\mathcal{I}_2$ of the (right) Cauchy--Green tensor of the deformation. In this expression, $C_{00}=0,C_{10},C_{01},C_{20},\dots$ are material parameters. The leading terms in this expansion constitute the Mooney--Rivlin density~\cite{rivlin1951,ogden,goriely}
\begin{subequations}
\begin{align}
W_1=C_{10}(\mathcal{I}_1-3)+C_{01}(\mathcal{I}_2-3),\label{eq:mr}
\end{align}
with thermodynamic stability, i.e., $W_1\ge 0$ for arbitrary deformations, requiring $C_{10},C_{01}\ge 0$~\cite{upadhyay19}. For a thin elastic shell of thickness $h$, this reduces to~\cite{audoly,ventsel,libai,gregory17,haas21}
\begin{align}
\hat{W}_1=4G_1h\left(E^2+\dfrac{h^2}{12}K^2\right),\label{eq:sl}
\end{align}
\end{subequations}
in which $E$ and $K$ are measures of the stretching and bending of the shell, and where ${G_1=C_{10}+C_{01}}$~\cite{dervaux09,haas21}. We will call this the linear shear modulus~\footnote{The Mooney--Rivlin energy density~\eqref{eq:mr} is itself highly nonlinear, but the corresponding shell theory~\eqref{eq:sl} is the same as that of a linearly elastic material. In the context of this work, we will therefore only refer to the higher-order nonlinearities expressed by Eq.~\eqref{eq:W2} as ``(purely) nonlinear elasticity'', because they do change the shell theory, as expressed by Eq.~\eqref{eq:snl}.}; it vanishes if and only if $C_{10}=C_{01}=0$. The terms in Eq.~\eqref{eq:W} expressing purely nonlinear elasticity are now led by
\begin{subequations}
\begin{align}
W_2=C_{20}(\mathcal{I}_1\!-\!3)^2+C_{11}(\mathcal{I}_1\!-\!3)(\mathcal{I}_2\!-\!3)+C_{02}(\mathcal{I}_2\!-\!3)^2.\label{eq:W2}
\end{align}
In the Supplemental Material~\footnote{See Supplemental Material at [url to be inserted], which includes Refs.~\cite{steigmann13,haas21,goriely,ogden,audoly,ventsel,libai,gregory17,dervaux09,pogorelov}, for (i) a derivation of the conditions for thermodynamic stability of $W_2$, (ii)~the derivation of the corresponding elastic shell theory and the nonlinear shear modulus $G_2$, (iii) the calculation of the prefactor for the Pogorelov scaling for a purely nonlinear elastic shell, (iv) additional finite-element simulation results, and (v) details of the derivation of the scaling diagrams for the force-displacement exponents of a pressurised and pre-stressed elastic shell.}, we show that $W_2$ is thermodynamically stable if and only if ${C_{20},C_{02}\ge 0}$ and $C_{11}\ge -2\sqrt{C_{20}C_{02}}$. \nocite{steigmann13,haas21,goriely,ogden,audoly,ventsel,libai,gregory17,dervaux09,pogorelov} We also show that, for axisymmetric deformations of a thin elastic shell, this becomes
\begin{align}
\hat{W}_2=16G_2h\left(E^4+\dfrac{h^2}{4}C^4+\dfrac{h^4}{80}K^4\right),\label{eq:snl}
\end{align}
\end{subequations}
where $C\sim\sqrt{EK}$ couples stretching and bending, and where we will refer to $G_2=C_{20}+C_{11}+C_{02}$ as the nonlinear shear modulus of the shell.

\begin{figure}[th!]
    \centering
    \includegraphics[width=\linewidth]{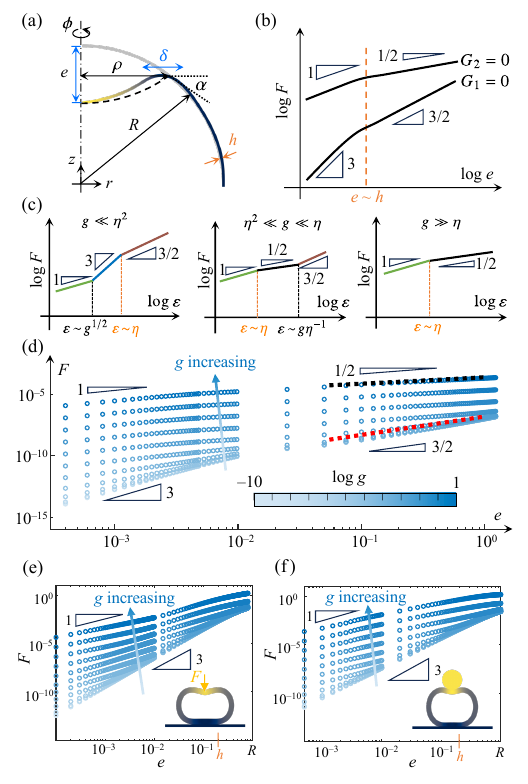}
    \caption{Nonlinear Pogorelov problem. (a)~Geometry: an indented spherical elastic shell of radius $R$ and thickness $h$ forms a dimple of depth $e$ and radius $\rho$ for $e\gg h$. The dimple ridge, over which the tangent angle changes from $-\alpha$ to $\alpha$, has extent $\delta$. (b)~Scaling exponents in the relation between indentation force $F$ and depth $e$: classical results for a linearly elastic shell ($G_2=0$) and results for a purely nonlinearly elastic shell ($G_1=0$). (c) Possible diagrams of scaling exponents in the force-displacement relationship, depending on the linearity $g=G_1/G_2$ and the non-dimensional thickness $\eta=h/R$. The critical non-dimensional indentations $\varepsilon=e/R$ for transitions between scaling exponents are indicated on the axes. (d) Force-displacement relation from finite-element simulations of the Pogorelov problem for a thin shell with $\eta=0.005$, for $g\in [10^{-10},10]$. Dotted lines: analytical approximations of the prefactor of the scaling relation for $e\gg h$, for $G_2=0$ (black) and $G_1=0$ (red). Parameter values satify $C_{10}=C_{01}$, $C_{20}=C_{11}=C_{02}$. (e)~Corresponding force-displacement relation for a thick shell ($\eta=0.2$) indented by a point force. (f)~Corresponding force-displacement relation for a thick shell indented by a sphere of radius $0.4R$.}
    \label{fig2}
\end{figure} 

We now solve the Pogorelov dimple problem for this purely nonlinear elasticity. The geometry of the problem is sketched in~\textfigref{fig2}{a}: A spherical shell of radius $R$ and thickness $h$ indented by an amount $e$ forms a dimple of radius $\rho$. For ${e\ll h}$, the elastic energy is concentrated in the dimple~\cite{landaulifshitz,sun2023geometric,Kierfeld2019shallow}, of area $A\sim \rho^2$, and geometry implies ${E\sim e/R}$, $K\sim e/\rho^2$. The shell energy is therefore $\smash{\mathcal{E}_2\sim A\hat{W}_2\sim G_2h e^4\bigl(\rho^2R^{-4}+h^2\rho^{-2}R^{-2}+h^4\rho^{-6}\bigr)}$, which is minimised for $\rho\sim(h R)^{1/2}$. For $e\gg h$, the elastic energy is concentrated at the rim of dimple instead~\cite{audoly,pogorelov,landaulifshitz,gomez16}, in a ridge of extent $\delta$ and area $A\sim\delta\rho$. This ridge joins the undeformed part of the shell to the isometrically inverted dimple, and the tangent angle changes from $\alpha$ to $-\alpha$ across the ridge~\figref{fig2}{a}. Geometry~\cite{audoly,pogorelov,landaulifshitz} now gives $\alpha\sim (e/R)^{1/2}$, $\rho\sim (eR)^{1/2}$ as well as ${E\sim \alpha\delta/R}$ and $K\sim\alpha/\delta$, so the shell energy is $\smash{\mathcal{E}_2\sim A\hat{W}_2\sim G_2he^{5/2}R^{-2}\bigl(\delta^5R^{-4}+h^2\delta R^{-2}+h^4\delta^{-3}\bigr)}$, minimised now for $\delta\sim (hR)^{1/2}$. The energy of a purely nonlinear Pogorelov dimple is therefore
\begin{align}
\mathcal{E}_2\sim\left\{\begin{array}{cl}
G_2e^{4}h^{2}R^{-3}&\text{for }e \ll h,  \\
G_2e^{5/2}h^{7/2}R^{-3}&\text{for }e \gg h.
\end{array}\right.\label{eq:E2}
\end{align}
Using $\hat{W}_1$ rather than $\hat{W}_2$, the same geometric scalings yield the classical energy of a linear Pogorelov dimple,
\begin{align}
\mathcal{E}_1\sim\left\{\begin{array}{cl}
G_1e^2h^2R^{-1}&\text{for }e \ll h,  \\
G_1e^{3/2}h^{5/2}R^{-1}&\text{for }e \gg h.
\end{array}\right.\label{eq:E1}
\end{align}
This energy balances the work $\mathcal{W}\sim Fe$ done by the indenting force. This gives new scaling exponents for a purely nonlinear Pogorelov dimple~\figref{fig2}{b}: $F\sim e^3$ for $e\ll h$, $F\sim e^{3/2}$ for $e\gg h$ if $G_1=0$, replacing the classical scalings $F\sim e$, $F\sim e^{1/2}$ for $G_2=0$.

For a general material with $G_1,G_2\neq 0$, we can approximate $\mathcal{E}\sim\mathcal{E}_1+\mathcal{E}_2$ because the dimple and ridge size scalings are the same for the classical and nonlinear Pogorelov problem. We introduce $g=G_1/G_2$ as a dimensionless measure of the linearity of the material, and the dimensionless shell thickness and indentation $\eta=h/R\ll 1$, $\varepsilon=e/R\ll 1$. With these definitions, $\mathcal{E}_1\gtrless\mathcal{E}_2$ if $\varepsilon\gtrless g^{1/2}$ and $\varepsilon\ll\eta$ or $\varepsilon\gtrless g/\eta$ and $\varepsilon\gg\eta$. Thus, if $g\gg\eta$, then $\mathcal{E}\sim\mathcal{E}_1$ for all $\varepsilon\ll 1$. If $\eta^2\ll g\ll\eta$, then $\mathcal{E}\sim\mathcal{E}_1$, $\mathcal{E}\sim\mathcal{E}_2$ for $\varepsilon\ll g/\eta,\varepsilon\gg g/\eta$, with $g/\eta\gg\eta$ in this case. Finally, if $g\ll\eta^2$, then $\mathcal{E}\sim\mathcal{E}_1$, $\mathcal{E}\sim\mathcal{E}_2$ for $\varepsilon\ll g^{1/2},\varepsilon\gg g^{1/2}$, where we note that $g^{1/2}\ll\eta$. This means that the force-displacement relation has three possible diagrams of scaling exponents~\figref{fig2}{c}. In particular, for $g\ll\eta$, nonlinear scaling exponents appear in the scaling diagram and can be used to quantify the nonlinearity of the material.

To verify these results, we performed finite-element simulations in \textsc{Comsol Multiphysics}, implementing the energy density $W_1+W_2$ in the ``Nonlinear Structural Materials Module''. The numerical results in \textfigref{fig2}{d} confirm the scaling exponents and the transitions between them. They also match the prefactor for the scaling relation~\eqref{eq:E1} and $e\gg h$, for which Pogorelov developed an analytical approximation~\cite{pogorelov,gomez16}, and the corresponding prefactor that we derive in the Supplemental Material~\cite{Note2} for the purely nonlinear scaling relation~\eqref{eq:E2} by extending Pogorelov's calculations.

The cysts that have motivated these calculations are not however thin shells and the microbeads used in indentation experiments~\figref{fig1}{a} do not exert point forces on them either. We therefore extend our numerical simulations to thick shells indented by a point force~\figref{fig2}{e} or a spherical indenter~\figref{fig2}{f}. Remarkably, the scaling exponents predicted by the asymptotic scalings for $e\ll h\ll R$ persist for these thick shells up to indentations comparable to $h$ and $R$. In the Supplemental Material~\cite{Note2}, we show further that these scalings also survive when a constraint fixing the lumen volume enclosed by the shell is imposed. All of this shows that our scaling relations~\figref{fig2}{c} can indeed represent the mechanical behaviour of cysts. Importantly, the Hertzian scaling $F\sim e^{3/2}$~\cite{Hertz,landaulifshitz,audoly} and the scaling $F\sim e^3$ of Ref.~\cite{Fery2004} do not show up in our simulations for $G_2=0$. This proves that, for the indentation of thick shells, ``large'' scaling exponents in the force-displacement relation must be ascribed to material nonlinearities.

The success of our scaling relations suggests that we can also understand more complex mechanical behaviour of cysts in this way. We therefore extend our scaling arguments to pressurised and pre-stressed cysts, focusing on the case $e\ll h$ because we have seen those scalings in our numerical simulations for thick shells. The work done by pressure $p$ is proportional to the volume $V\sim e^2R$ of the dimple, with corresponding energy
\begin{align}
\mathcal{E}_p\sim pe^2R.\label{eq:Ep}
\end{align}
Next, pre-stress due to cell contractility or adhesion induces a pre-strain $E_0$, which replaces ${E\to E+E_0}$ in Eqs.~\eqref{eq:sl} and~\eqref{eq:snl}. If ${E\gg E_0}$, the previous scalings continue to hold. If $E\ll E_0$, then, from Eq.~\eqref{eq:sl},
\begin{subequations}
\begin{align}
\mathcal{E}_1&\sim G_1h\left[R^2E_0^2+A\bigl(EE_0+h^2K^2\bigr)\right]\nonumber\\
&\sim G_1he\left(\dfrac{E_0\rho^2}{R}+\dfrac{h^2e}{\rho^2}\right)+\text{const.},
\end{align}
where we have reused the geometric scalings introduced above, and where the constant term is the integral of the pre-strain over the whole shell. This is minimised for $\rho\sim (eR/E_0)^{1/4}h^{1/2}$, so
\begin{align}
\mathcal{E}_1\sim \dfrac{G_1 h^2E_0^{1/2}e^{3/2}}{R^{1/2}} \quad\text{for $E\ll E_0$ (and $e\ll h$)}. \label{eq:E10}
\end{align}
\end{subequations}
Similarly, from Eq.~\eqref{eq:snl},
\begin{subequations}
\begin{align}
\mathcal{E}_2&\sim G_2h\left[R^2E_0^2+A\bigl(EE_0^3+h^4K^4+h^2E_0^2K^2\bigr)\right]\nonumber\\
&\sim G_2he\left(\dfrac{E_0^3\rho^2}{R}+\dfrac{h^4e^3}{\rho^6}+\dfrac{h^2eE_0^2}{\rho^2}\right)+\text{const.}.\label{eq:E2E0}
\end{align}
The first two terms (i.e., stretching and bending) are balanced for ${\rho\sim h^{1/2}R^{1/8}(e/E_0)^{3/8}}$. With this, we find $\bigl(h^2e E_0^2/\rho^2\bigr)\big/\bigl(E_0^3\rho^2/R\bigr)\sim (RE_0/e)^{1/2}\sim (E_0/E)^{1/2}\gg 1$, so the third term swamps the balanced first two terms, which cannot therefore minimise $\mathcal{E}_2$. Hence the minimising balance must be between the first and last terms in Eq.~\eqref{eq:E2E0}, i.e., between the stretching and coupling terms. This yields $\rho\sim h^{1/2}(eR/E_0)^{1/4}$ again, whence
\begin{align}
\mathcal{E}_2\sim\dfrac{G_2h^2E_0^{5/2}e^{3/2}}{R^{1/2}}\quad\text{for $E\ll E_0$ (and $e\ll h$)}. \label{eq:E20}
\end{align}
\end{subequations}
Now $\bigl(h^4e^3/\rho^6\bigr)\big/\bigl(E_0^3\rho^2/R\bigr)\sim e/E_0R\sim E/E_0\ll 1$, confirming that the second, bending term can indeed be neglected in this balance.

\begin{figure}[b]
    \centering
    \includegraphics{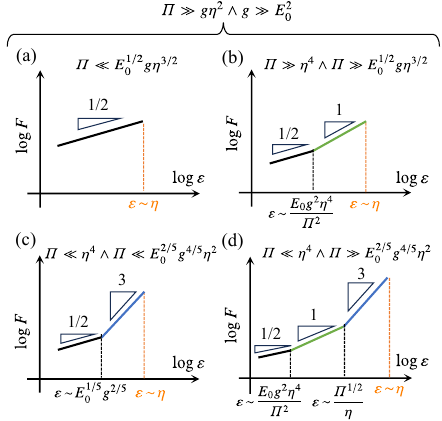}
    \caption{Diagrams of scaling exponents for a pressurized Pogorelov dimple with pre-strain $E_0$, shown for the strongly pressurised, weakly nonlinear case $\Pii \gg g \eta^2$, $g\gg E_0^2$, and for indentations $\varepsilon\ll\eta$. There are four different possible diagrams; the additional conditions in which each diagram occurs are shown above the figure panels, and the scalings of the critical indentations at which the exponents transition are indicated on the axes.}
    \label{fig3}
\end{figure}

Our results thus give three possible scaling exponents in the force-displacement relation for $e\ll h$: $1/2$, $1$, and~$3$, corresponding to the scalings $\mathcal{E}\sim e^{3/2}$, $\mathcal{E}\sim e^2$, $\mathcal{E}\sim e^4$. For small enough $e$, the first energy scaling must dominate, but, as $e$ increases, other scalings may take over and hence other, larger scaling exponents may appear one after the other. The stiffness ${\partial F/\partial e\sim\mathcal{E}/e^2}$ thus decreases for small $e$ (when the first scaling dominates), but can increase (i.e., there may be strain-stiffening) if the scaling $\mathcal{E}\sim e^4$ associated with nonlinear elasticity~[Eq.~\eqref{eq:E2}] dominates. There are thus \emph{a priori} four possible diagrams of scaling exponents~\wholefigref{fig3}. However, $\mathcal{E}\sim e^2$ is associated with linear elasticity or pressure [Eqs.~\eqref{eq:E1}, \eqref{eq:Ep}] if $G_1h^2R^{-1}\gtrless pR$, i.e., if $g\eta^2\gtrless\Pii$, where ${\Pii=p/G_2}$. Similarly, $\mathcal{E}\sim e^{3/2}$ is associated with pre-stressed linear or nonlinear elasticity~[Eqs.~\eqref{eq:E10}, \eqref{eq:E20}] if $\smash{G_1h^2E_0^{1/2}R^{-1/2}\gtrless G_2h^2E_0^{5/2}R^{-1/2}}$, i.e., if $\smash{g\gtrless E_0^2}$. The conditions under which these scaling diagrams arise therefore depend on whether $\Pii\gtrless g\eta^2$ and $E_0^2\gtrless g$.

In the Supplemental Material~\cite{Note2}, we derive the logical conditions for each of these scaling diagrams to occur, and show how to simplify them using \textsc{Mathematica} (Wolfram, Inc.). The resulting diagrams and conditions are shown in \textwholefigref{fig3} for the strongly pressurised, weakly nonlinear case $\Pii\gg g\eta^2$, $g\gg E_0^2$; we have relegated the other cases to the Supplemental Material~\cite{Note2}.

\begin{figure}[t]
    \centering
    \includegraphics{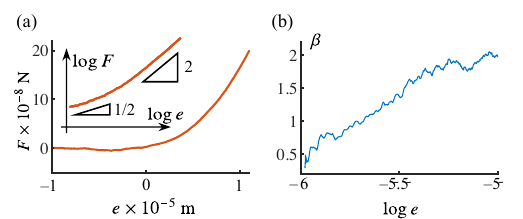}
    \caption{Observation of scaling exponents in cyst indentation experiments~\figref{fig1}{a}. (a) Example of force-displacement relation from indentation of a cyst of MDCK cells~\cite{Shen2017}. Experimental data provided by Pingbo Huang and Yusheng Shen. Inset: same data plotted on logarithmic axes, showing a scaling exponent increasing from $1/2$ to $2$. (b)~Plot of the local scaling exponent $\beta=\mathrm{d}(\log{F})/\mathrm{d}(\log{e})$ against $e$, estimated by smoothing the raw data in panel (a).}
    \label{fig4}
\end{figure}

We conclude by reanalysing experimental data on cyst indentation experiments~\cite{Shen2017} in the light of our results. The observed force-displacement relation~\figref{fig4}{a} reveals a scaling exponent $\beta$ increasing from $1/2$ to $2$ as the indentation depth is increased~\figrefi{fig4}{a}. The experimental resolution of a single indentation experiment does not of course allow quantifying ``true'' scaling exponents over many decades of indentation depth or visualising true transitions between exponents~\figref{fig4}{b}. Nonetheless, observing the exponent $\beta=1/2$ that we have predicted above emphasises the importance of pre-stress for the mechanical behaviour of the cyst. Experiments at deeper indentation would be required to determine whether the observed exponent $\beta=2$, also seen in other recent experiments~\cite{pereztirado24}, would indeed increase to the predicted exponent $\beta=3$. Nevertheless, observing strain-stiffening exponents $\beta>1$ in this way signals a contribution of nonlinear elasticity to the mechanics of the cyst.

In summary, we have combined analytical calculations, scaling arguments, and finite-element simulations to predict new scaling exponents in the force-displacement relation of an elastic shell from nonlinear elasticity and pre-stress. We have found signatures of these exponents in cyst indentation experiments. Our results therefore form the basis for interpreting the mechanical behaviour of cysts and other biological materials, although further experimental work will be needed to test the scalings and transitions between scalings regimes predicted here. Future work, both experimental and theoretical, will also need to explore further the effect on these scaling exponents of viscoelasticity~\cite{Andreas2020,pereztirado24} or superelasticity~\cite{Latorre2018,pereztirado24} for slow indentation on timescales comparable to those over which the cortex rearranges.

Meanwhile, deriving these scaling exponents more microscopically, for example from coarse-grained vertex models of cysts~\cite{bovyn24}, remains an important theoretical challenge. More generally, the purely nonlinear elasticity introduced here adds to the burgeoning field that seeks new hyperelastic energy densities to describe soft biological materials~\cite{martonova24,Kuhl2024}, opening up a host of mechanical questions: For example, how does the buckling of a shell under external pressure~\cite{knoche11,vella11,knoche14,Kierfeld2019shallow} change for such a ``supersoft'' material?

\begin{acknowledgments}
The authors are grateful to P. Huang and Y. Shen for making available the raw experimental data plotted in \textwholefigref{fig4}, and thank A. Janshoff, A. Honigmann, and A.~Perez-Tirado for discussions. This work was supported by the Max Planck Society; in particular, access to \textsc{Comsol Multiphysics} was provided by the Max Planck Computing and Data Facility.
\end{acknowledgments}

\bibliography{references}

\end{document}


\renewcommand{\theequation}{S\arabic{equation}}
\renewcommand{\thefigure}{S\arabic{figure}}
\renewcommand{\thepage}{S\arabic{page}}

\title{Mechanics of poking a cyst\\ \textbullet{} Supplemental Material \textbullet{}}

\author{Shiheng Zhao}
\affiliation{Max Planck Institute for the Physics of Complex Systems, N\"othnitzer Straße 38, 01187 Dresden, Germany}
\affiliation{Center for Systems Biology Dresden, Pfotenhauerstraße 108, 01307 Dresden, Germany}
\affiliation{\mbox{Max Planck Institute of Molecular Cell Biology and Genetics, Pfotenhauerstraße 108, 01307 Dresden, Germany}}
\author{Pierre A. Haas}
\affiliation{Max Planck Institute for the Physics of Complex Systems, N\"othnitzer Straße 38, 01187 Dresden, Germany}
\affiliation{Center for Systems Biology Dresden, Pfotenhauerstraße 108, 01307 Dresden, Germany}
\affiliation{\mbox{Max Planck Institute of Molecular Cell Biology and Genetics, Pfotenhauerstraße 108, 01307 Dresden, Germany}}
\maketitle

This Supplemental Material is divided into five parts. In the first part, we derive the conditions for thermodynamic stability of the purely nonlinear strain energy function. In the second part, we derive the corresponding elastic shell theory, which we use in the third part to solve the Pogorelov problem for such a purely nonlinear elastic shell. We then present further finite-element simulation results. In the final part, we derive the scaling diagrams for the force-displacement exponents of a pressurised and pre-stressed elastic shell.

\section{Thermodynamic stability of purely nonlinear elasticity}
Thermodynamic stability requires that the purely nonlinear strain energy function introduced in the main text be non-negative, i.e.,
\begin{align}
W=C_{20}(\mathcal{I}_1-3)^2+C_{11}(\mathcal{I}_1-3)(\mathcal{I}_2-3)+C_{02}(\mathcal{I}_2-3)^2\geq 0\label{eq:sse}
\end{align}
for all deformations, where $C_{20},C_{11},C_{02}$ are material parameters, and $\mathcal{I}_1,\mathcal{I}_2$ are the principal invariants of the left Cauchy-Green tensor. To analyse this condition, we first express $\mathcal{I}_1,\mathcal{I}_2$ in terms of the principal stretches $\Lambda_1,\Lambda_2,\Lambda_3$. By incompressibility, assumed here, $\Lambda_3=1/\Lambda_1\Lambda_2$, so
\begin{align}
    &\mathcal{I}_1 = \Lambda_1^2+\Lambda_2^2+\frac{1}{\Lambda_1^2\Lambda_2^2},&&
    \mathcal{I}_2 = \Lambda_1^2\Lambda_2^2+\frac{1}{\Lambda_1^2}+\frac{1}{\Lambda_2^2},
\end{align}
From the inequality of arithmetic and geometric means, we find
\begin{align}
    \mathcal{I}_1 &\ge3\sqrt[3]{\Lambda_1^2\Lambda_2^2\frac{1}{\Lambda_1^2\Lambda_2^2}}=3,&\mathcal{I}_2 &\geq3\sqrt[3]{\Lambda_1^2\Lambda_2^2\frac{1}{\Lambda_1^2}\frac{1}{\Lambda_2^2}}=3,
\end{align}
with equality if and only if $\Lambda_1=\Lambda_2=1$. This proves that the ratio $R=(\mathcal{I}_2-3)/(\mathcal{I}_1-3)$ is a positive, continuous function of $\Lambda_1,\Lambda_2>0$, with a removable singularity at $\Lambda_1=\Lambda_2=1$. If $\Lambda_1=\Lambda_2=\Lambda\gg 1$, then $R\sim \Lambda^2/2\gg 1$. If $\Lambda_1=\Lambda^{-1/2}$ and $\Lambda_2=\Lambda\gg 1$, then $R\sim2/\Lambda\ll 1$. By continuity, this shows that $R$ attains all positive real values for $\Lambda_1,\Lambda_2>0$. This means that
\begin{equation}
\dfrac{W}{(\mathcal{I}_1-3)^2}=C_{02}R^2+C_{11}R+C_{20}\geq 0\quad\text{for all }R>0.\label{eq:sscond}
\end{equation}
If $C_{02}=0$, it is clear that condition~\eqref{eq:sscond} holds if and only if $C_{11}\geq 0$ and $C_{20}\geq 0$. If $C_{02}\neq0$, considering the limits $R\gg 1$ and $R\ll 1$ respectively, it is clear that it holds only if $C_{02}\geq0$ and $C_{20}\geq 0$. Conversely, if $C_{20}\geq 0$ and $C_{02}>0$, then the quadratic in $R$ on the left-hand side of the inequality is non-negative for sufficiently small and sufficiently large $R>0$, so condition~\eqref{eq:sscond} breaks down if and only if it has two positive real roots. The product of its roots is $C_{20}/C_{02}>0$, so if the roots are real, which is if and only if $\smash{C_{11}^2-4C_{20}C_{02}}\geq 0$, they are both positive if and only if their sum $-C_{11}/C_{02}$ is positive, which is now if and only if $C_{11}<0$. Hence, if $C_{02}\neq 0$, condition~\eqref{eq:sscond} holds if and only if $C_{02}>0$ and $C_{20}\geq 0$ and $\smash{C_{11}^2<4C_{20}C_{02}}$ or $C_{11}>0$. Combining all of these conditions shows that the purely nonlinear strain energy function~\eqref{eq:sse} is non-negative if and only if
\begin{equation}
   C_{20}\geq0\quad\text{and}\quad C_{02}\geq 0\quad\text{and}\quad C_{11}\geq -2\sqrt{C_{20}C_{02}}.
\end{equation}

\section{Purely nonlinear elastic shell theory}
In this section, we derive a theory for purely nonlinear elastic shells, following prior approaches for linearly elastic shells~\cite{steigmann13,haas21}, with our notation based on that of Ref.~\cite{haas21}.

We restrict to torsionless deformations of an axisymmetric shell. The thickness of the shell is $\varepsilon h$, where $\varepsilon\ll 1$ is an asymptotically smal parameter that describes the thinness of the shell compared to other lengthscales associated with its midsurface. We begin by deriving the elastic deformation gradient $\tens{F}$ of the shell.

\begin{figure}[t]
    \centering
    \includegraphics{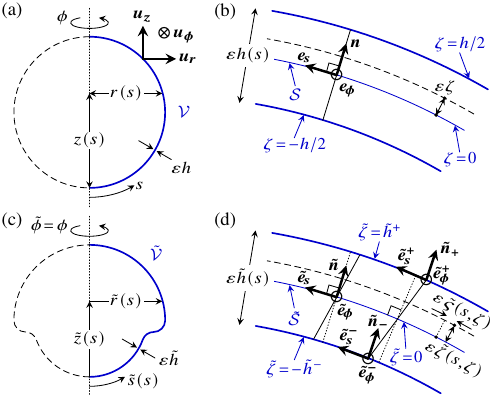}
    \caption{Deformations of an axisymmetric shell. (a)~Undeformed configuration $\mathcal{V}$ of an axisymmetric shell of thickness $\varepsilon h(s)$, defined by the coordinates $r(s),z(s)$, where $s$ is arclength, with respect to the basis $\{\vec{u_r},\vec{u_\phi},\vec{u_z}\}$ of cylindrical polars. (b)~Cross-section of the undeformed shell, showing the basis $\mathcal{B}=\{\vec{e_s},\vec{e_\phi},\vec{n}\}$ of the midsurface $\mathcal{S}$ and the transverse coordinate~$\zeta$. The surfaces of the undeformed shell are at $\zeta=\pm h/2$. (c)~Deformed configuration~$\smash{\tilde{\mathcal{V}}}$ of the shell: After a torsionless deformation, the thickness of the shell is $\varepsilon\tilde{h}(s)$, its arclength is $\tilde{s}(s)$, and its coordinates with respect to cylindrical polars are $\tilde{r}(s),\tilde{z}(s)$. (d)~Cross section of the deformed shell, showing the basis $\tilde{\mathcal{B}}=\{\vec{\tilde{e}_s},\vec{\tilde{e}_\phi},\vec{\tilde{n}}\}$ of its midsurface $\tilde{\mathcal{S}}$. Due to rotation of the normals to the midsurface, a point at a distance $\varepsilon\zeta$ from $\mathcal{S}$ ends up at at a $\varepsilon\tilde{\zeta}(s,\zeta)$ from $\tilde{\mathcal{S}}$, and displaced by an amount $\varepsilon\tilde{\varsigma}(s,\zeta)$ parallel to $\tilde{\mathcal{S}}$. The surfaces of the deformed shell are at $\tilde{\zeta}=\pm\tilde{h}^\pm(s)$, where the tangent vectors are $\vec{\tilde{e}_s^\pm},\vec{\tilde{e}_{\smash\phi}^\pm}$ and the normal is $\vec{\tilde{n}^\pm}$. Figure redrawn from Ref.~\cite{haas21}.}
    \label{figs1}
\end{figure}

\subsection{Calculation of the elastic deformation gradient}
To derive the deformation gradient tensor that maps the initial (undeformed) configuration $\mathcal{V}$ of the shell to its current deformed configuration $\tilde{\mathcal{V}}$, we review the geometry and kinematics of the axisymmetric shell.

We describe the undeformed configuration $\mathcal{V}$ of the shell with reference to its midsurface $\mathcal{S}$. With respect to the basis $\{\vec{u_r},\vec{u_\phi},\vec{u_z}\}$ of cylindrical coordinates, the position vector of a point on $\mathcal{S}$ is
\begin{align}
\vec{\rho}(s,\phi)=r(s)\vec{u_r}(\phi)+z(s)\vec{u_z},\label{eq:mundef}
\end{align}
where $s$ is arclength along the curve generating the axisymmetric midsurface, and $\phi$ is the azimuthal coordinate, so that $r(s)$ and $z(s)$ are the radial and vertical positions of a point on the midsurface~\figref{figs1}{a}. The tangent angle $\psi(s)$ of $\mathcal{S}$ is defined by
\begin{align}
&r'(s)=\cos{\psi(s)},\quad z'(s)=\sin{\psi(s)},  
\end{align}
in which dashes denote differentiation with respect to $s$. Thus
\begin{align}
&\vec{e_s}(s,\phi)=\cos{\psi(s)}\vec{u_r}(\phi)+\sin{\psi(s)}\vec{u_z},\quad \vec{e_\phi}(\phi)=\vec{u_\phi}(\phi) 
\end{align}
provide a basis of the tangent space of $\mathcal{S}$, from which we obtain a right-handed orthonormal basis ${\mathcal{B}=\{\vec{e_s},\vec{e_\phi},\vec{n}\}}$ for $\mathcal{V}$ by adjoining the normal to $\mathcal{S}$,
\begin{align}
\vec{n}(s,\phi)=\cos{\psi(s)}\vec{u_z}-\sin{\psi(s)}\vec{u_r}(\phi).
\end{align}
The curvatures of $\mathcal{S}$ are found to be
\begin{align}
&\varkappa_s(s)=\psi'(s),&&\varkappa_\phi(s)=\dfrac{\sin{\psi(s)}}{r(s)}.\label{eq:curvundef}
\end{align}
This completes the description of the midsurface $\mathcal{S}$. The position of a point in $\mathcal{V}$ is now
\begin{align}
\vec{r}(s,\phi,\zeta)=\vec{\rho}(s,\phi)+\varepsilon\zeta\vec{n}(s,\phi), \label{eq:undef}
\end{align}
where $\zeta$ is the transverse coordinate such that the midsurface is at $\zeta=0$ and that, by the definition, the shell surfaces are at $\zeta=\pm h/2$. Since $\partial\vec{n}/\partial s=-\varkappa_s\vec{e_s}$ and $\partial\vec{n}/\partial\phi=-\varkappa_\phi\vec{e_\phi}$, the tangent basis of $\mathcal{V}$ is
\begin{align}
&\dfrac{\partial\vec{r}}{\partial s}=(1-\varepsilon\varkappa_s\zeta)\vec{e_s},\quad \dfrac{\partial\vec{r}}{\partial \phi}=r(1-\varepsilon\varkappa_\phi\zeta)\vec{e_\phi},\quad \dfrac{\partial\vec{r}}{\partial\zeta}=\varepsilon\vec{n}.\label{eq:pdundef}
\end{align}
The scale factors of the undeformed configuration are therefore\begin{subequations}
\begin{align}
&\chi_s=1-\varepsilon\varkappa_s\zeta,&& \chi_\phi=r(1-\varepsilon\varkappa_\phi\zeta),&&\chi_\zeta=\varepsilon.\label{eq:sfundef}
\end{align}
and the volume element of the undeformed shell is
\begin{align}
\mathrm{d}V=\chi_s\chi_\phi \chi_\zeta\,\mathrm{d}s\,\mathrm{d}\phi\,\mathrm{d}\zeta=\varepsilon(1-\varepsilon\varkappa_s\zeta)(1-\varepsilon\varkappa_\phi\zeta)\,r\,\mathrm{d}s\,\mathrm{d}\phi\,\mathrm{d}\zeta.\label{eq:dVundef}
\end{align}
\end{subequations}
As the shell deforms into its deformed configuration $\tilde{\mathcal{V}}$, the undeformed midsurface $\mathcal{S}$ maps to the deformed midsurface $\tilde{\mathcal{S}}$, which has position vector
\begin{align}
\vec{\tilde{\rho}}(s,\phi)=\tilde{r}(s)\vec{u_r}(\phi)+\tilde{z}(s)\vec{u_z}, 
\end{align}
defining the positions $\tilde{r}(s),\tilde{z}(s)$ of a point on $\tilde{\mathcal{S}}$~\figref{figs1}{c}. Let $\tilde{s}$ denote the deformed arclength. The stretches of the deformed shell are thus
\begin{align}
&f_s(s)=\dfrac{\mathrm{d}\tilde{s}}{\mathrm{d}s},&&f_\phi(s)=\dfrac{\tilde{r}(s)}{r(s)},\label{eq:fs}
\end{align}
and so the tangent angle $\tilde{\psi}(s)$ of $\tilde{\mathcal{S}}$ is defined by
\begin{align}
&\tilde{r}'(s)=f_s\cos{\tilde{\psi}(s)},\quad \tilde{z}'(s)=f_s\sin{\tilde{\psi}(s)}.  \label{eq:drz}
\end{align}
As for the undeformed configuration, we introduce the tangent vectors
\begin{align}
&\vec{\tilde{e}_s}(s,\phi)=\cos{\tilde{\psi}(s)}\vec{u_r}(\phi)+\sin{\tilde{\psi}(s)}\vec{u_z},\quad\vec{\tilde{e}_\phi}(\phi)=\vec{u_\phi}(\phi), 
\end{align}
and the normal vector
\begin{align}
\vec{\tilde{n}}(s,\phi)=\cos{\tilde{\psi}(s)}\vec{u_z}-\sin{\tilde{\psi}(s)}\vec{u_r}(\phi), 
\end{align}
They define a right-handed orthonormal basis ${\tilde{\mathcal{B}}=\{\vec{\tilde{e}_s},\vec{\tilde{e}_\phi},\vec{\tilde{n}}\}}$ of \smash{$\tilde{\mathcal{V}}$}. The curvatures of the deformed shell are
\begin{align}
&\kappa_s(s)=\dfrac{\tilde{\psi}'(s)}{f_s(s)},&&\kappa_\phi(s)=\dfrac{\sin{\tilde{\psi}(s)}}{\tilde{r}(s)}.\label{eq:kappas}
\end{align}
Under the mapping from $\mathcal{V}$ to $\tilde{\mathcal{V}}$, normals to $\mathcal{S}$ will not in general map to normals to $\tilde{\mathcal{S}}$. We will therefore assume that a point in $\mathcal{V}$ at a distance $\varepsilon\zeta$ from $\mathcal{S}$ is mapped to a point at a distance $\varepsilon\tilde{\zeta}$ from and displaced by a distance $\varepsilon\tilde{\varsigma}$ parallel to $\tilde{\mathcal{S}}$~\figref{figs1}{d}. The definition of the midsurfaces implies that $\tilde{\zeta}=\tilde{\varsigma}=0$ if $\zeta=0$. A point in $\tilde{\mathcal{V}}$ thus has position vector
\begin{align}
\vec{\tilde{r}}(s,\phi,\zeta)= \vec{\tilde{\rho}}(s,\phi)+\varepsilon\tilde{\zeta}(s,\zeta)\vec{\tilde{n}}(s,\phi)+\varepsilon\tilde{\varsigma}(s,\zeta)\vec{\tilde{e}_s}(s,\phi). \label{eq:def}
\end{align}
We now compute
\begin{subequations}\label{eq:pddef}
\begin{align}
\dfrac{\partial\vec{\tilde{r}}}{\partial s}&=\left[\tilde{f}_s\left(1-\varepsilon\tilde{\kappa}_s \tilde{\zeta}\right)+\varepsilon \tilde{\varsigma}_{,s}\right]\vec{\tilde{e}_s}+\varepsilon\left(\tilde{\zeta}_{,s}+\tilde{f}_s\tilde{\kappa}_s \tilde{\varsigma}\right)\vec{\tilde{n}},\label{eq:pddefs}\\
\dfrac{\partial\vec{\tilde{r}}}{\partial\phi}&=\left[\tilde{r}\left(1-\varepsilon\tilde{\kappa}_\phi \tilde{\zeta}\right)+\varepsilon \tilde{\varsigma}\cos{\tilde{\psi}}\right]\vec{\tilde{e}_\phi},\\
\dfrac{\partial\vec{\tilde{r}}}{\partial\zeta}&=\varepsilon\left(\tilde{\zeta}_{,\zeta}\vec{\tilde{n}}+\tilde{\varsigma}_{,\zeta}\vec{\tilde{e}_s}\right).
\end{align}
\end{subequations}
By definition~\cite{goriely}, the deformation gradient tensor is
\begin{align}
\tens{F}=\operatorname{Grad}{\vec{\tilde{r}}}=\dfrac{1}{\chi_s^2}\dfrac{\partial\vec{\tilde{r}}}{\partial s}\otimes\dfrac{\partial\vec{r}}{\partial s}+\dfrac{1}{\chi_\phi^2}\dfrac{\partial\vec{\tilde{r}}}{\partial\phi}\otimes\dfrac{\partial\vec{r}}{\partial\phi}+\dfrac{1}{\chi_\zeta^2}\dfrac{\partial\vec{\tilde{r}}}{\partial\zeta}\otimes\dfrac{\partial\vec{r}}{\partial\zeta},
\end{align}
and we thus obtain, with respect to $\tilde{\mathcal{B}}\otimes\mathcal{B}$,
\begin{align}
\tens{F}&=\left(\begin{array}{ccc}
\dfrac{f_s\left(1\!-\!\varepsilon\kappa_s \tilde{\zeta}\right)\!+\!\varepsilon\tilde{\varsigma}_{,s}}{1-\varepsilon\varkappa_s\zeta}&0&\tilde{\varsigma}_{,\zeta}\\
0&\dfrac{f_\phi\left(1\!-\!\varepsilon\kappa_\phi \tilde{\zeta}\right)\!+\!\dfrac{\varepsilon\tilde{\varsigma}}{r}\cos{\tilde{\psi}}}{1-\varepsilon\varkappa_\phi\zeta}&0\\
\dfrac{\varepsilon\left(\tilde{\zeta}_{,s}+f_s\kappa_s\tilde{\varsigma}\right)}{1-\varepsilon\varkappa_s\zeta}&0&\tilde{\zeta}_{,\zeta}
\end{array}\right).
\label{eq:F}
\end{align}

\subsection{Elastic energy and scaling assumptions}
In this subsection, we derive the effective elastic energy for a ``supersoft'' shell by asymptotic expansion of three-dimensional elasticity. Let $\tens{C}=\tens{F}^\top\tens{F}$ denote the left Cauchy--Green tensor, and let $\mathcal{I}_1=\tr{\tens{C}}$ and $\mathcal{I}_2=\bigl(\mathcal{I}_1^2-\tr{\tens{C}^2}\bigr)/2$ be its invariants~\cite{goriely,ogden}. The elastic energy is
\begin{subequations}
\begin{align}
\mathcal{E}=\iiint_{\mathcal{V}}{W\,\mathrm{d}V},\label{eq:E}
\end{align}
where $W$ is the purely nonlinear energy density
\begin{align}
W=C_{20}(\mathcal{I}_1-3)^2+C_{11}(\mathcal{I}_1-3)(\mathcal{I}_2-3)+C_{02}(\mathcal{I}_2-3)^2.\label{eq:E2}
\end{align}
\end{subequations}
In Eq.~\eqref{eq:E}, the integration is over the initial configuration $\mathcal{V}$ of the shell, and $\mathrm{d}V$ is the volume element given by Eq.~\eqref{eq:dVundef}. 

From its general expression~\cite{goriely}, the Cauchy stress tensor associated with this energy density is seen to be
\begin{align}
\tens{T}&=2\bigl[(\mathcal{I}_1-3)(2C_{\text{20}}+C_{\text{11}}\mathcal{I}_1)+(\mathcal{I}_2-3)(C_{\text{11}}+2C_{\text{02}} \mathcal{I}_1)\bigr]\tens{B} \nonumber\\
    &\quad-2\bigl[C_{\text{11}} (\mathcal{I}_1-3)+2 C_{\text{02}} (\mathcal{I}_2-3)\bigr]\tens{B}^2-p\tens{I},
\end{align}
in which $\tens{I}$ is the identity, $\tens{B}=\tens{F}\tens{F}^\top$ is the right Cauchy--Green tensor, and the pressure $p$ is the Lagrange multiplier that imposes the condition $\det{\tens{F}}=1$ of incompressibility. The associated first Piola--Kirchhoff stress tensor $\tens{P} =\tens{T} \tens{F}^{-\top}$ satisfies the equilibrium equation~\cite{ogden}
\begin{subequations}
\begin{align}
    \text{Div}~\tens{P}^{\top} = \vec{0}, 
\end{align}
where the divergence is with respect to the undeformed configuration. As pointed out in Ref.~\cite{haas21}, since $\mathcal{B},\tilde{\mathcal{B}}$ are independent of $\zeta$, the equilibrium equation can separated into components perpendicular and parallel to the midsurface, viz., 
\begin{align}
    \frac{(\tens{P}\vec{n})_{,\zeta}}{\varepsilon} + \vec{\nabla}\cdot \tens{P}^\top = \vec{0},
    \label{eq:Pbalance}
\end{align}
\end{subequations}
where $\vec{\nabla}$ denotes the gradient on $\mathcal{S}$. The force-free boundary conditions on the surfaces of the deformed shell are $\tens{T}^{\vec{\pm}}\vec{\tilde{n}^\pm}=\vec{0}$, where $\vec{\tilde{n}^\pm}$ are the normals to $\tilde{\zeta}=\tilde{h}^\pm$~\figref{figs1}{d}, and $\tens{T}^{\vec{\pm}}$ are the values of the Cauchy stress there. These conditions are well known to be equivalent to $\tens{P}^{\vec{\pm}}\vec{n^\pm}=\vec{0}$~\cite{ogden}, where, since the undeformed shell is assumed to have constant thickness $h$, $\vec{n^\pm}=\vec{n}$ are the normals to $\zeta=\pm h/2$, and $\tens{P}^{\vec{\pm}}$ are the values of the first Piola--Kirchhoff stress there. 
\subsubsection*{\textbf{Scaling assumptions}}
To derive the shell theory, we need to expand the elastic energy~\eqref{eq:E} in the asymptotic limit $\varepsilon\ll 1$. For this purpose, we need to make scaling assumptions for the stretches and curvatures of the deformed shell,
\begin{align}
f_s =1+\varepsilon E_s +O\bigl(\varepsilon^2\bigr),\quad f_\phi =1+\varepsilon E_\phi+O\bigl(\varepsilon^2\bigr),\label{eq:shellstr}
\end{align}
and
\begin{align}
\kappa_s=\varkappa_s+K_s+O(\varepsilon),\quad \kappa_\phi=\varkappa_\phi+K_\phi+O(\varepsilon).\label{eq:curvstr}
\end{align}
These relations define the shell strains $E_s,E_\phi$ and the curvature strain $K_s,K_\phi$.

\subsection{Asymptotic expansion}
To obtain an asymptotic expansion of the elastic energy subject to the scalings assumptions above, we posit regular expansions
\begin{align}
\tilde{\zeta}=\tilde{\zeta}_{(0)}+\varepsilon \tilde{\zeta}_{(1)}+O\bigl(\varepsilon^2\bigr),\quad \varsigma=\varsigma_{(0)}+O(\varepsilon),
\label{eq:expansions}
\end{align}
for the transverse and parallel displacements. We further expand
\begin{subequations}
\begin{align}
\tens{P}&=\tens{P}_{(0)}+\varepsilon\tens{P}_{(1)}+\varepsilon^2\tens{P}_{(2)}+\varepsilon^3\tens{P}_{(3)}+O\bigl(\varepsilon^4\bigr),\\
\label{eq:Qpexpansions}
p&=p_{(0)}+p_{(1)}\varepsilon+p_{(2)}\varepsilon^2+O(\varepsilon^3).
\end{align}
\end{subequations}
\subsubsection*{\textbf{Asymptotic expansion of the incompressibility condition}}
We begin by expanding the incompressibility condition to leading-order using Eq.~\eqref{eq:F} to get
\begin{align}
1=\det{\tens{F}}=\tilde{\zeta}_{(0),\zeta}+O(\varepsilon),
\label{eq:det0}
\end{align}
Integrating this result using $\tilde{\zeta}_{(0)}=0$ at $\zeta=0$, as required by the definition of the midsurface, yields $\smash{\tilde{\zeta}_{(0)}}\equiv \zeta$. We now expand the incompressibility condition further, finding
\begin{align}
0=\det{\tens{F}}-1=\varepsilon\left(E_s+E_\phi-\zeta (K_\phi+K_s) +\dfrac{\partial \tilde{\zeta}_{(1)}}{\partial \zeta}\right)+O\bigl(\varepsilon^2\bigr). 
\label{eq:det1}
\end{align}
This provides a differential equation for $\tilde{\zeta}_{(1)}$, the solution of which that satisfies $\tilde{\zeta}_{(1)}=0$ at $\zeta=0$ is
\begin{align}
\tilde{\zeta}_{(1)}=\dfrac{K_s+K_\phi}{2}\zeta^2-(E_s+E_\phi)\zeta.
\label{eq:Z1}
\end{align}
\begin{widetext}
\subsubsection*{\textbf{Asymptotic expansion of the constitutive relations}}
It turns out that we do not need to expand the deformation gradient explicitly beyond order~$O(\varepsilon)$. Rather, it suffices to consider a formal expansion
\begin{align}
\tens{F}=\left(\begin{array}{ccc}
1+\varepsilon a_{(1)}+\varepsilon^2 a_{(2)}+\varepsilon^3 a_{(3)}+O\bigl(\varepsilon^4\bigr)&0&\varepsilon v_{(1)}+\varepsilon^2 v_{(2)}+\varepsilon^3 v_{(3)}+O\bigl(\varepsilon^4\bigr)\\
0&1+\varepsilon b_{(1)}+\varepsilon^2 b_{(2)}+\varepsilon^3 b_{(3)}+O\bigl(\varepsilon^4\bigr)&0\\
\varepsilon w_{(1)}+\varepsilon^2 w_{(2)}+\varepsilon^3 w_{(3)}+O\bigl(\varepsilon^4\bigr)&0&1+\varepsilon c_{(1)}+\varepsilon^2 c_{(2)}+\varepsilon^3 c_{(3)}+O\bigl(\varepsilon^4\bigr)
\end{array}\right),\label{eq:F2}
\end{align}
where, by comparison with Eq.~\eqref{eq:F} and using the previous results,
\begin{align}
a_{(1)}=E_s-\zeta K_s,\quad b_{(1)} =E_\phi-\zeta K_\phi.\label{eq:a1b1}
\end{align}
Expressions for $a_{(2)},a_{(3)},b_{(2)},b_{(3)},c_{(1)},c_{(2)},c_{(3)},v_{(1)},v_{(2)},w_{(1)},w_{(2)}$ could of course also be obtained from Eq.~\eqref{eq:F} in terms of the expansions~\eqref{eq:expansions}, but will not be needed. The incompressibility condition is now
\begin{align}
1=\det{\tens{F}}&=1+\varepsilon\left(a_{(1)}+b_{(1)}+c_{(1)}\right)+\varepsilon^2\left(a_{(2)}+b_{(2)}+c_{(2)}+a_{(1)}b_{(1)}+b_{(1)}c_{(1)}+c_{(1)}a_{(1)}-v_{(1)}w_{(1)}\right)+O\bigl(\varepsilon^3\bigr).
\label{eq:detFexp}
\end{align}
This yields, successively,
\begin{align}
c_{(1)}&=-a_{(1)}-b_{(1)},&c_{(2)}&=a_{(1)}^2+a_{(1)}b_{(1)}+b_{(1)}^2-a_{(2)}-b_{(2)}+v_{(1)}w_{(1)}.\label{eq:BCres}
\end{align}
Now, at leading order, Eq.~\eqref{eq:Pbalance} yields $(\tens{P}_{(0)}\vec{n})_{, \zeta}=\vec{0}$, which shows that $\tens{P}_{(0)}\vec{n}$ is independent of $\zeta$. The boundary conditions require $\vec{0} = \tens{P}^{\pm}\vec{n}=\tens{P}_{(0)}\vec{n}+O(\varepsilon)$. This shows that $\tens{P}_{(0)}\vec{n}\equiv\vec{0}$. Direct calculation using \textsc{Mathematica} (Wolfram, Inc.) reveals that this implies that $p_{(0)}=0$, and hence $\tens{P}_{(0)}\equiv\mathbfsf{0}$. The equilibrium condition~\eqref{eq:Pbalance} now yields $(\tens{P}_{(1)}\vec{n})_{, \zeta}=0$, and a similar argument leads to $p_{(1)}=0$ and $\tens{P}_{(1)}\equiv\mathbfsf{0}$. In a similar manner, we obtain $\tens{P}_{(2)}\vec{n}\equiv\vec{0}$, whence, by direct calculation,
\begin{align}
p_{(2)}=2(4C_{\text{02}}+3C_{\text{11}}+2C_{\text{20}})\bigl[4a_{(1)}^2+4a_{(1)}b_{(1)}+4b_{(1)}^2+(v_{(1)}+w_{(1)})^2\bigr],    
\end{align}
and thus $\tens{P}_{(2)}\equiv\mathbfsf{0}$. In turn and in the now familiar fashion, we then obtain $\tens{P}_{(3)}\vec{n}\equiv\vec{0}$. Now
\begin{align}
\tens{P}_{(3)}\vec{n}=\left(\begin{array}{c}
4(C_{20}+C_{11}+C_{02})(v_{(1)}+w_{(1)})\bigl[2a_{(1)}^2+2(a_{(1)}+b_{(1)})^2+2b_{(1)}^2+(v_{(1)}+w_{(1)})^2\bigr]\\
0\\
O(1)
\end{array}\right).
\end{align}
It is clear that the factor in square brackets only vanishes if all the terms in it vanish; in particular, its vanishing requires $v_{(1)}+w_{(1)}=0$. If the factor in square brackets does not vanish, then $\tens{P}_{(3)}\vec{n}\equiv\vec{0}$ still requires $v_{(1)}+w_{(1)}=0$, assuming that the nonlinear shear modulus $G_2\equiv C_{20}+C_{11}+C_{02}$ does not vanish (a possibility that we are not interested in). Hence
\begin{align}
w_{(1)}=-v_{(1)}\label{eq:BCres2}
\end{align}
in either case. On computing the expansion of $\tens{C}=\tens{F}^\top\tens{F}$ from Eq.~\eqref{eq:F} and hence those of $\mathcal{I}_1=\tr{\tens{C}}$ and $\mathcal{I}_2=\bigl(\mathcal{I}_1^2-\tr{\tens{C}^2}\bigr)/2$, and simplifying using Eqs.~\eqref{eq:BCres} and \eqref{eq:BCres2}, we then obtain
\begin{subequations}
\begin{align}
\mathcal{I}_1&=3+\varepsilon\bigl[2\bigl(a_{(1)}+b_{(1)}+c_{(1)}\bigr)\bigr]+\varepsilon^2\left[a_{\smash{(1)}}^2+b_{\smash{(1)}}^2+c_{\smash{(1)}}^2+v_{\smash{(1)}}^2+w_{\smash{(1)}}^2+2\left(a_{(2)}+b_{(2)}+c_{(2)}\right)\right]+O\bigl(\varepsilon^3\bigr)\nonumber\\
&=3+\varepsilon^2\bigl[4\bigl(a_{\smash{(1)}}^2+a_{(1)}b_{(1)}+b_{\smash{(1)}}^2\bigr)\bigr]+O\bigl(\varepsilon^3\bigr)=\mathcal{I}_2+O\bigl(\varepsilon^3\bigr).
\label{eq:I1}
\end{align}
Hence, from Eqs.~\eqref{eq:a1b1},
\begin{align}
\mathcal{I}_1-3&=4 \varepsilon^2 \left[(E_s-\zeta K_s)(E_\phi-\zeta K_\phi)+(E_s-\zeta K_s)^2+(E_\phi-\zeta K_\phi)^2\right]+O\bigl(\varepsilon^3\bigr)=\mathcal{I}_2-3.
\label{eq:I2}
\end{align}
\end{subequations}
This determines the leading-order term in the asymptotic expansion of the energy density in Eq.~\eqref{eq:E2}, and thus allows us to derive the shell energy.
\subsection{Derivation of the shell energy}
Substituting Eq.~\eqref{eq:I2} into Eq.~\eqref{eq:E2}, integrating with respect to $\zeta$, and using axisymmetry, the integral expression~\eqref{eq:E} for the elastic energy finally reduces to
\begin{subequations}\label{eq:st}
\begin{align}
\mathcal{E}=\iint_{\mathcal{S}}{\hat{W}\,r\,\mathrm{d}s\,\mathrm{d}\phi}=2\pi\int_{\mathcal{C}}{\hat{W}\,r\,\mathrm{d}s}, \label{eq:Ea}
\end{align}
where the first integral is over the undeformed axisymmetric midsurface $\mathcal{S}$ and the second one is over the curve $\mathcal{C}$ generating $\mathcal{S}$. In Eq.~\eqref{eq:Ea}, $\hat{W}$ is the effective two-dimensional energy density,
\begin{align}
\hat{W}=\varepsilon\int_{-h/2}^{h/2}{W\,\mathrm{d}\zeta}&=\varepsilon ^5 G_2\left\{16h\bigl(E_s^2+E_s E_{\phi }+E_{\phi}^2\bigr)^2+\frac{h^5}{5}\bigl(K_s^2+K_s K_{\phi}+K_{\phi }^2\bigr)^2\right.\nonumber\\
    &\qquad\qquad\quad+4h^3\left[E_s^2K_s^2+(E_s+E_\phi)^2(K_s+K_{\phi})^2+E_{\phi }^2K_{\phi }^2\right]
    \biggr\}+O\bigl(\varepsilon^6\bigr),
\label{eq:edens}
\end{align}
\end{subequations}
\end{widetext}
where $G_2=C_{20}+C_{11}+C_{02}$ is the nonlinear shear modulus, as defined in the main text. We obtain the non-asymptotic form of this shell theory by setting $\varepsilon=1$ or, more formally, by replacing ${\varepsilon h\to h}$, $\varepsilon E_s\to E_s$, and $\varepsilon E_\phi\to E_\phi$. This result is seen to be of the form in Eq.~(3b) of the main text by identifying the strain measures used there with
\begin{subequations}
\begin{align}
E&=\bigl(E_s^2+E_sE_\phi+E_\phi^2\bigr)^{1/2},\\
K&=\bigl(K_s^2+K_sK_\phi+K_\phi^2\bigr)^{1/2},\\
C&=\bigl[E_s^2K_s^2+(E_s+E_\phi)^2(K_s+K_{\phi})^2+E_{\phi }^2K_{\phi }^2\bigr]^{1/4}.
\end{align}
\end{subequations}
In particular, $C\sim\sqrt{EK}$. Now the shell energy density for a Mooney--Rivlin material is known to be~\cite{audoly,ventsel,libai,gregory17,haas21}
\begin{align}
4G_1h\left[E_s^2+E_sE_\phi+E_\phi^2+\dfrac{h^2}{12}\bigl(K_s^2+K_sK_\phi+K_\phi^2\bigr)\right],
\end{align}
with $G_1=C_{10}+C_{01}$~\cite{dervaux09,haas21}, so these definitions also recover Eq.~(2b) of the main text.
\section{Calculation of the prefactor of the force-displacement relation}
\begin{figure}[b]
    \centering
    \includegraphics[width=\linewidth]{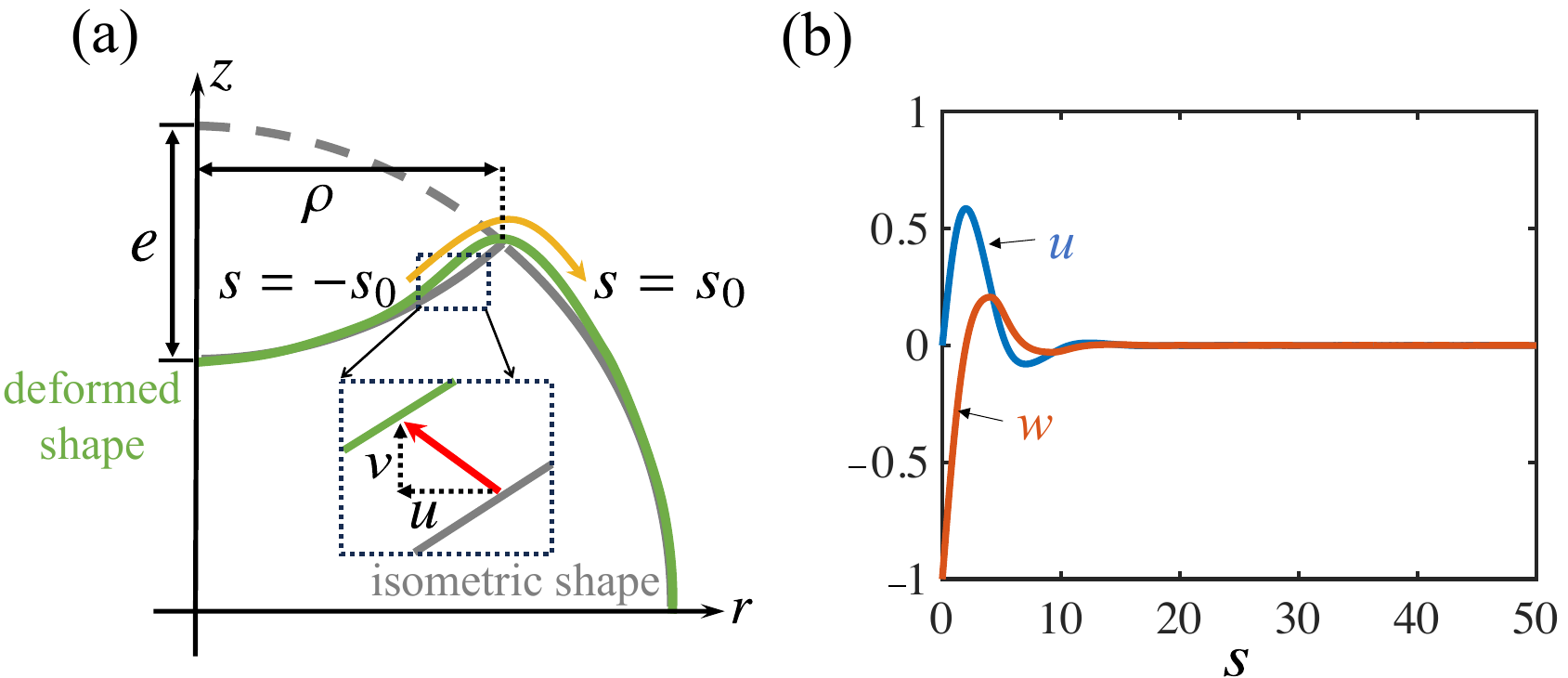}
    \caption{Calculation of the prefactor in the force-displacement relation for indentation of a ``supersoft'' shell. (a)~Geometry of the problem: The deformed shape of the shell differs from the isometric dimple shape by small displacements $u(s),v(s)$ in the radial and axial directions of cylindrical polars $(r,z)$, where $s$ is arclength, with the ridge area corresponding to $-s_0\leq s\leq s_0$. The indentation of the dimple is $e$. (b) Numerical solution of Eqs.~\eqref{eq:elvar}.}
    \label{figs2}
\end{figure}

Just as Pogorelov extended the arguments that determine the scaling of the force with the indentation to a calculation of the corresponding prefactor~\cite{pogorelov}, we can extend the scaling arguments of the main text to calculate the prefactors. Following Pogorelov's approach~\cite{pogorelov}, we describe the deformed shape of the shell by its displacements $u(s)$ and $v(s)$, in the radial and axial directions of cylindrical polar coordinates respectively, from the isometric dimple shape~\figref{figs2}{a}, where $s$ is arclength along the isometric shape, with $s=0$ corresponding to its ridge. 

For a thin shell, the elastic energy is concentrated in the ridge of the dimple, where Pogorelov estimated~\cite{pogorelov}
\begin{align}
&E_\phi\approx\dfrac{u}{\rho}\gg E_s,&&K_s \approx v'' \gg K_\phi,
\end{align}
in which $\rho$ is the radius of the dimple~\figref{figs2}{a}. The elastic energy of a ``supersoft'' dimpled shell is therefore, from Eqs.~\eqref{eq:st},
\begin{align}
\mathcal{E} &\approx G_2\iint_{\mathcal{R}}{\left[16h\left(\dfrac{u}{\rho}\right)^4+4h^3\left(\dfrac{u}{\rho}\right)^2v''^2+\dfrac{h^5}{5}v''^4\right]\mathrm{d}S},
\end{align}
where the integration region $\mathcal{R}$ is the ridge of the dimple, $-s_0\leq s\leq s_0$, and the surface element is $\mathrm{d}S\approx 2\pi \rho\,\mathrm{d}s$. Across the ridge, the tangent angle to the shell changes from $-\alpha$ to $\alpha$, with $\rho=R\sin{\alpha}\approx R\alpha$. We introduce scaled variables
\begin{align}
    s &= \xi \overline{s}, & u &=\xi\alpha^2\overline{u}, &&\dfrac{\mathrm{d}v}{\mathrm{d}s}=\alpha\overline{w}\;\Rightarrow\;\dfrac{\mathrm{d}^2v}{\mathrm{d}s^2}=\frac{\alpha}{\xi}\frac{\mathrm{d}\overline{w}}{\mathrm{d}\overline{s}},
\end{align}
where we choose
\begin{align}
    \xi \equiv \frac{\alpha^{-1/2}h^{1/2}\rho^{1/2}}{80^{1/8}}.
\end{align}
The energy of the shell becomes
\begin{equation}
\mathcal{E}=2a\int_{0}^{\overline{s}_0}{\left(\overline{u}^4+\overline{w}'^4 +\sqrt{5}\overline{u}^2\overline{w}'^2 \right)\mathrm{d}\overline{s}},\label{eq:Edimple}
\end{equation}
where $\overline{s}_0=s_0/\xi$ is the dimensionless arclength of the ridge and where
\begin{equation}
a=\frac{32\pi G_2}{80^{5/8}}h^{7/2}\rho^{-1/2}\alpha^{11/2}\approx \frac{32\pi G_2}{80^{5/8}}h^{7/2}e^{5/2}R^{-3},
\end{equation}
in terms of the indentation ${e=2(R-R\cos{\alpha})\approx R\alpha^2}$ of the dimple, using $\alpha\approx (e/R)^{1/2}$, $\rho\approx (eR)^{1/2}$.

Using $\overline{s}_0\gg 1$ to take the upper limit of integration in Eq.~\eqref{eq:Edimple} to infinity and dropping overlines, we are left to determine 
\begin{align}
J=\min{\int_{0}^{\infty}{\left(u^4+w'^4+\sqrt{5}u^2w'^2 \right)\mathrm{d}s}},    
\end{align}
subject to geometric boundary conditions that carry over from Pogorelov's calculation~\cite{pogorelov},
\begin{subequations}
\begin{align}
&u(0) = u(\infty) = 0,&&w(0) = -1, &&w(\infty) = 0,
\label{eq:bounvar}
\end{align}
and subject to the compatibility condition~\cite{pogorelov}
\begin{equation}
u'+w+ \frac{w^2}{2}= 0.
\label{eq:compvar}
\end{equation}    
\end{subequations}
To solve this functional minimisation problem, we introduce the Lagrangian
\begin{align}
\mathcal{L}=\int_{0}^{\infty}{\left[u^4+w'^4+\sqrt{5}u^2w'^2-\lambda\left(u'+w+ \frac{w^2}{2}\right)\right]\mathrm{d}s},
\end{align}
where $\lambda(s)$ is a Lagrange multiplier function that imposes the constraint~\eqref{eq:compvar}. We find the resulting Euler--Lagrange equations to be
\begin{subequations}\label{eq:elvar}
\begin{align}
u'&=-w-\dfrac{w^2}{2},\\
w''&=-\dfrac{4\sqrt{5}uu'w'+\lambda(1+w)}{12w'^2+2\sqrt{5}u^2},\\
\lambda'&=-4u^3-2\sqrt{5}uw'^2.
\end{align}
\end{subequations}
We solve this fourth-order system of highly nonlinear differential equations subject to the four boundary conditions~\eqref{eq:bounvar} using \textsc{Comsol Multiphysics}. The solution for $u,w$ is plotted in~\textfigref{figs2}{b}, and by numerical integration, we find $J_2\approx0.56$. The force-displacement relation is now obtained by balancing the work
\begin{align}
\mathcal{W}=\int_0^e{F(e)\,\mathrm{d}e}    
\end{align}
done by the indentation force $F$ with the elastic energy $\mathcal{E}$. Equivalently, $F=\partial\mathcal{E}/\partial e$. We thus find
\begin{equation}
F = \left(\frac{160\pi J_2}{80^{5/8}}\right)G_2h^{7/2}R^{-3}e^{3/2}\approx 18.1G_2h^{7/2}R^{-3}e^{3/2}.
\end{equation}

\section{Additional finite-element simulation results}
\begin{figure}[t]
    \centering
    \includegraphics{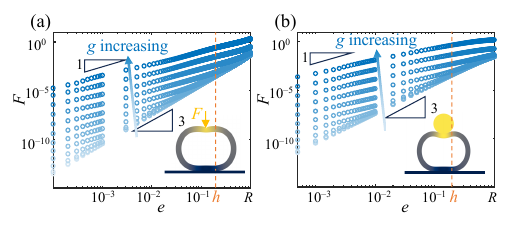}
    \caption{Finite-element simulations of a thick shell ($h/R=0.2$) indented with (a) a point load or (b) a sphere of radius $0.4R$, subject to the constraint of lumen volume conservation.}
    \label{figs3}
\end{figure}

In this section, we add the constraint of lumen volume conservation to our finite-element simulations of cyst indentation. We keep all geometrical and material parameters consistent with Figs.~2(e),(f) of the main text. The results in \textwholefigref{figs3} show that volume conservation does not have a significant effect on the scaling exponents of the force-displacement relation, neither for a point load~\figref{figs3}{a} nor for a spherical indenter~\figref{figs3}{b}. ``Large'' scaling exponents are only seen in the nonlinear elastic case $g\ll 1$. As discussed in the main text, these results show that the scaling exponent $3$ is associated, for indentation of thick cysts, with material nonlinearities rather than volume conservation.

\section{Force-displacement diagrams for the pressurised, pre-stressed shells}
As in the main text, we consider the force-displacement relation for a pressurised Pogorelov dimple with pre-strain $E_0$. Let $R$ be the radius of the shell, and let $h$ be its thickness. We denote by $p$ be pressure of the dimple, and by $G_1$ and $G_2$ the linear and nonlinear shear moduli of the shell. The relevant non-dimensional parameters, also used in the main text, are thus $\eta=h/R$, $\Pi=p/G_2$, $g=G_1/G_2$.

We start by defining the conditions for scaling transitions. Let $e$ denote the indentation of the shell, and let $F$ denote the corresponding indentation force. Again as in the main text, we focus on the case $\varepsilon\equiv e/h\ll 1$. The balance between energies with scaling exponent $i$ and $j$ leads to the critical value $\varepsilon =\varepsilon^\ast_{i\to j}$, at which
\begin{align}
    \mathcal{E}_i(\varepsilon_{i\to j}^\ast) \sim \mathcal{E}_j(\varepsilon_{i\to j}^\ast),
\end{align}
and hence the scaling transition can happen. It does happen if and only if $\mathcal{E}_i(\varepsilon_{i\to j}^\ast) \sim \mathcal{E}_j(\varepsilon_{i\to j}^\ast)$ swamp all other energy contributions at this critical value, i.e., if and only if the condition
\begin{align}\label{scalingcond}
    \mathcal{C}_{i\to j}\colon\quad\mathcal{E}_i(\varepsilon_{i\to j}^\ast) \sim \mathcal{E}_j(\varepsilon_{i\to j}^\ast) \gg \mathcal{E}_k(\varepsilon_{i\to j}^\ast),~ \forall k\neq i, j
\end{align}
holds. As in the main text, we note that the scaling exponents can only increase. In particular, for sufficiently small $\varepsilon$, the exponent will be $1/2$. 

\begin{figure*}[t]
    \centering
    \includegraphics[width=\linewidth]{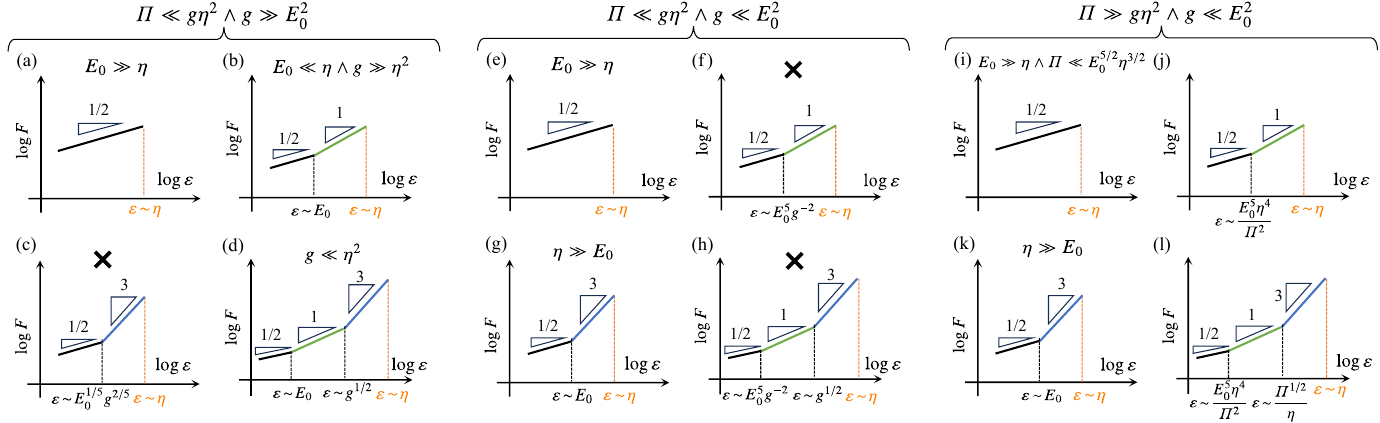}
    \caption{Pressurised Pogorelov dimple problem with pre-strain $E_0$, for an elastic shell of radius $R$ and thickness $h$ with pressure $p$, linear shear modulus $G_1$, and nonlinear shear modulus $G_2$: scaling diagrams of the force $F$ against indentation $e$, in terms of the non-dimensional parameters $\eta=h/R$, $\varepsilon=e/h$, $\Pi=p/G_2$, $g=G_1/G_2$, and for indentations $\varepsilon\ll\eta$. Diagrams are shown for the cases $\Pi\ll g\eta^2$, $\smash{g\gg E_{\smash{0}}^2}$ [panels (a)--(d)], $\Pi \ll g\eta^2$, $\smash{g\ll E_0^2}$ [panels (e)--(h)], and $\Pi \gg g\eta^2$, $\smash{g\ll E_0^2}$ [panels (i)--(l)]; the diagrams for the remaining case $\Pi\gg g\eta^2$, $\smash{g\gg E_0^2}$ are shown in Fig.~3 of the main text. The additional conditions in which each diagram occurs are shown above each figure panel, except for panels (j) and (l) for which the conditions are more complicated, and the scalings of the critical indentations at which the exponents transition are indicated on their axes. The condition for panel (j) is $\smash{\bigl[E_0\ll \eta \land \bigl(\Pi \ll E_{\smash{0}}^{10}g^{-4}\eta^{2} \lor \Pi\gg \eta^4\bigr)\bigr]\lor\bigl(E_0\gg\eta \land \Pi \gg E_{\smash{0}}^{5/2}\eta^{3/2}\bigr)}$, that for panel (l) is $E_0^2\eta^2\ll\Pi\ll\varepsilon^4\land\Pi\gg E_0^{10}g^{-4}\eta^2$. The symbol $\times$ is used to marked scaling diagrams that cannot arise in some of the three cases.}
    \label{figs4}
\end{figure*}

\subsection*{Scaling diagram conditions}
We first express the energy scalings determined in the main text in terms of the non-dimensional variables $\eta,\Pi,g,\varepsilon,E_0$. We will use subscripts to indicate the resulting scaling exponents in the force-displacement diagrams. Thus
\begin{subequations}
\begin{align}
\mathcal{E}_{3}&\sim \eta^2\varepsilon^4,
\end{align}
corresponding to the case of the mechanics being dominated by purely nonlinear elasticity. The other exponents are more interesting: The discussion in the main text shows that a linear-force displacement relation arises both if linear elasticity dominates [Eq.~(4) of the main text, $\mathcal{E}\sim g\eta^2\varepsilon^2$] and if pressure dominates [Eq.~(6) of the main text, $\mathcal{E}\sim \Pi\varepsilon^2$]. Similarly, in the pre-stressed case, the same scaling exponent of $1/2$ arises if linear elasticity dominates [Eq.~(7b) of the main text, $\smash{\mathcal{E}\sim gE_{\smash{0}}^{1/2}\eta^2\varepsilon^{3/2}}$] and if nonlinear elasticity dominates [Eq.~(8b) of the main text, $\smash{\mathcal{E}\sim E_{\smash{0}}^{5/2}\eta^2\varepsilon^{3/2}}$]. Thus,
\begin{align}
\mathcal{E}_1\sim\left\{\begin{array}{cl}
g\eta^2\varepsilon^2&\text{if }\Pi\ll g\eta^2,\\
\Pi\varepsilon^2&\text{if }\Pi\gg g\eta^2,
\end{array}\right.
\end{align}
and
\begin{align}
\mathcal{E}_{1/2}\sim\left\{\begin{array}{cl}
E_0^{5/2}\eta^2\varepsilon^{3/2}    &\text{if }g \ll E_0^2,\\
gE_0^{1/2}\eta^2\varepsilon^{3/2}   &\text{if }g \gg E_0^2.
\end{array}\right.
\end{align}
\end{subequations}
We shall refer to $\Pi\ll g\eta^2$ as the weakly pressurised case and $\Pi\gg g\eta^2$ as the strongly pressurised case. Similarly, $g\gg E_0^2$ is the weakly nonlinear case and $g\ll E_0^2$ is the strongly nonlinear case.

To determine the scaling diagrams, we need to determine the critical indentations $\smash{\varepsilon^\ast_{\smash{i\to j}}}$ at which $\smash{\mathcal{E}_i(\varepsilon_{\smash{i\to j}}^\ast) \sim \mathcal{E}_j(\varepsilon_{\smash{i\to j}}^\ast)}$. The above shows that there are four cases: 
\begin{center}
(1) $\Pi\ll g\eta^2$, $g\gg E_0^2$
\end{center}
\begin{subequations}
\begin{align}
\varepsilon_{1/2\to 1}^\ast&\sim E_0&&\Longleftarrow\; gE_0^{1/2}\eta^2\varepsilon^{3/2}\sim g\eta^2\varepsilon^2,\\
\varepsilon_{1/2\to 3}^\ast&\sim E_0^{1/5}g^{2/5}&&\Longleftarrow\; gE_0^{1/2}\eta^2\varepsilon^{3/2}\sim \eta^2\varepsilon^4,\\
\varepsilon_{1\to 3}^\ast&\sim g^{1/2}&&\Longleftarrow\; g\eta^2\varepsilon^2 \sim \eta^2\varepsilon^4;
\end{align}
\end{subequations}
\begin{center}
(2) $\Pi\ll g\eta^2$, $g\ll E_0^2$
\end{center}
\begin{subequations}
\begin{align}
\varepsilon_{1/2\to 1}^\ast&\sim E_0^5g^{-2} &&\Longleftarrow\;
E_0^{5/2}\eta^2\varepsilon^{3/2} \sim g\eta^2\varepsilon^2,\\
\varepsilon_{1/2\to 3}^\ast&\sim E_0 &&\Longleftarrow\;
E_0^{5/2}\eta^2\varepsilon^{3/2} \sim \eta^2\varepsilon^4,\\
\varepsilon_{1\to 3}^\ast&\sim g^{1/2}&&\Longleftarrow\; 
g\eta^2\varepsilon^2 \sim \eta^2\varepsilon^4;
\end{align}
\end{subequations}

\begin{center}
(3) $\Pi\gg g\eta^2$, $g\gg E_0^2$
\end{center}
\begin{subequations}
\begin{align}
\varepsilon_{1/2\to 1}^\ast&\sim E_0g^2\eta^4\Pi^{-2} &&\Longleftarrow\;gE_0^{1/2}\eta^2\varepsilon^{3/2} \sim \Pi\varepsilon^2,\\
\varepsilon_{1/2\to 3}^\ast&\sim E_0^{1/5}g^{2/5}&&\Longleftarrow\; gE_0^{1/2}\eta^2\varepsilon^{3/2}\sim \eta^2\varepsilon^4,\\
\varepsilon_{1\to 3}^\ast&\sim \Pi^{1/2}\eta^{-1}&&\Longleftarrow\; \Pi\varepsilon^2 \sim \eta^2\varepsilon^4;
\end{align}
\end{subequations}
\begin{center}
(4) $\Pi\gg g\eta^2$, $g\ll E_0^2$
\end{center}
\vspace{-10pt}
\begin{subequations}
\begin{align}
\varepsilon_{1/2\to 1}^\ast&\sim E_0^5\eta^4\Pi^{-2}&&\Longleftarrow\;
E_0^{5/2}\eta^2\varepsilon^{3/2} \sim g\eta^2\varepsilon^2,\\
\varepsilon_{1/2\to 3}^\ast&\sim E_0&&\Longleftarrow\;
E_0^{5/2}\eta^2\varepsilon^{3/2} \sim \eta^2\varepsilon^4,\\
\varepsilon_{1\to 3}^\ast&\sim \Pi^{1/2}\eta^{-1}&&\Longleftarrow\; g\eta^2\varepsilon^2 \sim \eta^2\varepsilon^4.
\end{align}
\end{subequations}
Using \textsc{Mathematica}, we can thence compute the conditions introduced in Eq.~\eqref{scalingcond}. Since the scaling exponents must increase with $\varepsilon$, there are four scaling diagrams that can arise \emph{a priori}. The conditions in which these diagrams arise are
\begin{subequations}\label{eq:conds}
\begin{align}
    \text{only}~1/2&: ~(\varepsilon_{1/2\to1}^\ast\gg\eta) ~\land~ (\varepsilon_{1/2\to3}^\ast\gg\eta),\\
    1/2\to 1&:~(\varepsilon_{1/2\to1}^\ast\ll\eta)
    ~\land~\mathcal{C}_{1/2\to1}~\land~ \\ \nonumber
    &\qquad (\neg \mathcal{C}_{1\to3} ~\lor~ \varepsilon_{1\to3}^\ast\gg\eta),\\
    1/2 \to 3&:~(\varepsilon_{1/2\to3}^\ast\ll\eta) ~\land~\mathcal{C}_{1/2\to3},\\
    1/2\to 1\to 3&:~(\varepsilon_{1\to3}^\ast\ll\eta)~\land~\mathcal{C}_{1/2\to1}~\land~\mathcal{C}_{1\to3}.
\end{align}    
\end{subequations}
We simplify these logical conditions using the \texttt{Reduce} function of \textsc{Mathematica} to determine which diagrams are possible and the conditions in which the possible diagrams do arise, in each of the four cases above. We check these computations by verifying in particular that the union of the resulting conditions is true. The results are shown in Fig.~3 of the main text for the strongly pressurised, weakly nonlinear case $\Pi\gg g\eta^2$, $g\gg E_0^2$, and in Fig.~\ref{figs4} for the remaining cases.

\bibliography{references}